\documentclass [11pt,a4paper] {article}

\usepackage[cp1252]{inputenc}
\usepackage[english]{babel}

\usepackage{amssymb}
\usepackage{amsmath}
\usepackage{amsfonts,amssymb}

\usepackage[dvips]{graphicx}

\DeclareMathAlphabet{\mathpzc}{OT1}{pzc}{m}{it}

\begin{document}

\title{Can category-theoretic semantics resolve the problem of the interpretation of the quantum state vector?}

\author{Arkady Bolotin\footnote{$Email: arkadyv@bgu.ac.il$} \\ \textit{Ben-Gurion University of the Negev, Beersheba (Israel)}}

\maketitle

\begin{abstract}\noindent Do correctness and completeness of quantum mechanics jointly imply that quantum state vectors are necessarily in one-to-one correspondence with elements of the physical reality? In terms of category theory, such a correspondence would stand for an isomorphism, so the problem of the status of the quantum state vector could be turned into the question of whether state vectors are necessarily isomorphic to elements of the reality.\\

\noindent As it is argued in the present paper, in order to tackle this question, one needs to complement the category-theoretic approach to quantum mechanics with the computational-complexity-theoretic considerations. 
Based on such considerations, it is demonstrated in the paper that the hypothesis of the isomorphism existing between state vectors and elements of the reality is expected to be unsuitable for a generic quantum system.\\

\noindent \textbf{Keywords:} Quantum state vector status, Category theory, Isomorphism, Computational-complexity theory, Identity morphism, Quantum state tomography, Quantum adiabatic algorithm.\\
\end{abstract}

\section{Introduction}

\noindent According to the famous argument made in the EPR paper \cite{Einstein}, every element of the physical reality must have a counterpart in the \textit{complete} physical theory. On the other hand, every prediction (i.e., an element) of the \textit{correct} physical theory must correspond to an element of the physical reality. Thus, it seems natural to assume that the correct and complete physical theory must be \textit{equivalent} to the physical reality in such a way that every element of the theory must be in one-to-one correspondence with an element of the reality.\\

\noindent Let us denote elements of a physical theory as $\mathpzc{X}$ and elements of the physical reality as $\mathpzc{Y}$. Then, using the language of category theory (in a spirit loosely related to the categorical approach to theories of physical systems \cite{Doering11,Baez} or to the program of categorical quantum mechanics \cite{Abramsky,Gogioso,Blass}), the equivalence between the theory and the reality can be expressed as an \textit{isomorphism} $\mathpzc{Y}\simeq\mathpzc{X}$, i.e., a one-to-one relation (morphism $f$) assigning to each element of the reality $\mathpzc{Y}$ an element of the theory $\mathpzc{X}$\\

\begin{equation} \label{1} 
   \begin{array}{cl}
      f:
      &
      \mathpzc{Y}  \rightarrow \mathpzc{X}
      \;\;\;\; 
   \end{array}  
\end{equation}
\smallskip

\noindent that can be “undone” in the sense that there is another one-to-one relation (morphism $g$)\\

\begin{equation} \label{2} 
   \begin{array}{cl}
      g:
      &
      \mathpzc{X}  \rightarrow \mathpzc{Y}
      \;\;\;\;  ,
   \end{array}  
\end{equation}
\smallskip

\noindent which plays role of the inverse of $f$. If those morphisms $f$ and $g$ are considered as “transformations” (between elements of the reality $\mathpzc{Y}$ and the theory $\mathpzc{X}$), then the composition $g\circ\! f\!\!:  \mathpzc{Y}\rightarrow\mathpzc{Y}$ will be the identity morphism $\mathrm{id}_\mathpzc{Y}$ for the elements of reality $\mathpzc{Y}$ and correspondingly the composition $f\!\circ\! g\!:  \mathpzc{X}\rightarrow\mathpzc{X}$ will be the identity morphism $\mathrm{id}_\mathpzc{X}$ for the elements of the theory $\mathpzc{X}$.\\

\noindent For example, in classical mechanics (as well as in thermodynamics and statistical mechanics) the `direct' morphism $f$ can denote a one-to-one relation that assigns to each physical state $\mathpzc{Y}$ characterized by the set of values $\lbrace\mathpzc{y}_1,…,\mathpzc{y}_N \rbrace$ of all possible degrees of freedom or parameters of a mechanical system a point in the multidimensional space $\mathpzc{X}=(\mathpzc{x}_1,…,\mathpzc{x}_N )$ (the phase space) so that the system's state $\mathpzc{Y}(t)$ evolving over time $t$ would trace a path $\mathpzc{X}(t)$ through this phase space:\\

\begin{equation} \label{3} 
   \begin{array}{cl}
      f:
      &
      \mathpzc{Y}\!\left(t\right)  \rightarrow \mathpzc{X}\!\left(t\right)
      \;\;\;\;  .
   \end{array}  
\end{equation}
\smallskip

\noindent Let the morphism $\mathcal{T}_t:  \mathpzc{X}(0)\rightarrow\mathpzc{X}(t)$ be a transformation of elements on the phase space $\mathpzc{X}$ (e.g., a propagator operating on the phase space) that describes a deterministic and reversible time evolution $\mathpzc{Y}(0)\rightarrow\mathpzc{Y}(t)$ of the mechanical system. Then, the composition of the `inverse' morphism $g$ and the morphism $\mathcal{T}_t$ would imply a one-to-one correspondence between the system's predicted state $\mathpzc{X}(t)$ compatible with the certain initial condition $\mathpzc{X}(0)$ and the system's physical state $\mathpzc{Y}(t)$\\

\begin{equation} \label{4} 
   \begin{array}{cl}
      g\circ\! \mathcal{T}_t:
      &
      \mathpzc{X}\!\left(0\right)
               \stackrel{\;\; \mathcal{T}_t}{{-\!\!\!\longrightarrow}} 
      \mathpzc{X}\!\left(t\right)
               \stackrel{\;\; g}{{-\!\!\!\longrightarrow}}
      \mathpzc{Y}\!\left(t\right)
      \;\;\;\;  .
   \end{array}  
\end{equation}
\smallskip

\noindent In the case of quantum mechanics, the direct morphism $f$ can be a one-to-one relation assigning to each set of values of physical quantities $\mathpzc{Y}=\lbrace\mathpzc{y}_1,…,\mathpzc{y}_N \rbrace$ characterizing the physical state (or \textit{the ontic state} to use the terminology introduced in \cite{Harrigan}) of a quantum system a unit vector  $\left|\!\left.{\Psi} \!\right.\right\rangle$ (a state vector) residing in a complex separable Hilbert space (whose exact nature is dependent on the system):\\

\begin{equation} \label{5} 
   \begin{array}{cl}
      f:
      &
      \mathpzc{Y}\!\left(t\right)  \rightarrow \left|\!\left.{\Psi}\!\left(t\right)\!\right.\right\rangle
      \;\;\;\;  .
   \end{array}  
\end{equation}
\smallskip

\noindent Let the automorphism $\mathcal{T}_t: \left|\!\left.{\Psi}\!\left(0\right)\!\right.\right\rangle \!\!\rightarrow\!\! \left|\!\left.{\Psi}\!\left(t\right)\!\right.\right\rangle$ corresponding to the reversible time evolution $\mathpzc{Y}\!\left(0\right) \!\rightarrow\!\mathpzc{Y}\!\left(t\right)$ of the quantum system be determined by the unitary transformation (unitary operator) $U\!\left(H(t),0,t\right)$ on the Hilbert space of state vectors $\left|\!\left.{\Psi} \!\right.\right\rangle$\\

\begin{equation} \label{6} 
      U\!\left(H(t),0,t\right)\left|\!\left.{\Psi}\!\left(0\right)\!\right.\right\rangle
      =
     \left|\!\left.{\Psi}\!\left(t\right)\!\right.\right\rangle
      \;\;\;\;   
\end{equation}
\smallskip

\noindent such that for small $\delta t$ the unitary operator $U\!\left(H(t),0,\delta t\right)$ would take the form\\

\begin{equation} \label{7} 
      U\!\left(H(t),0,\delta t\right)
      =
       \mathrm{id}^{1}_{\left|\!\left.{\Psi} \!\right.\right\rangle}
      - \frac{i}{\hbar}H(t)\delta t
      \;\;\;\;  ,
\end{equation}
\smallskip

\noindent where $H(t)$ is the system's Hamiltonian and $\mathrm{id}^{1}_{\left|\!\left.{\Psi} \!\right.\right\rangle}\!=\!f\!\circ\! g$ is the identity morphism (identity transformation) on the Hilbert space of the system's state vectors $\left|\!\left.{\Psi} \!\right.\right\rangle$. Then the composition $g\!\circ\! \mathcal{T}_t$ would entail the set of morphisms that for a range of time arguments $t\geq 0$ maps the system's predicted quantum states $\left|\!\left.{\Psi}\!\left(t\right)\!\right.\right\rangle$ compatible with the given initial state $\left|\!\left.{\Psi}\!\left(0\right)\!\right.\right\rangle$ to the system's ontic states $\mathpzc{Y}\!\left(t\right)$\\

\begin{equation} \label{8} 
   \begin{array}{cl}
      g\circ\! \mathcal{T}_t:
      &
      \left|\!\left.{\Psi}\!\left(0\right)\!\right.\right\rangle
               \stackrel{\;\; \mathcal{T}_t}{{-\!\!\!\longrightarrow}} 
      \left|\!\left.{\Psi}\!\left(t\right)\!\right.\right\rangle 
               \stackrel{\;\; g}{{-\!\!\!\longrightarrow}} 
      \mathpzc{Y}\!\left(t\right)
      \;\;\;\;  .
   \end{array}  
\end{equation}
\smallskip

\noindent As long as the reverse automorphism $\mathcal{T}_{-t}: \left|\!\left.{\Psi}\!\left(t\right)\!\right.\right\rangle \!\!\rightarrow\!\! \left|\!\left.{\Psi}\!\left(0\right)\!\right.\right\rangle$ is determined through $U\!\left(H(t),t,0\right)\left|\!\left.{\Psi}\!\left(t\right)\!\right.\right\rangle=\left|\!\left.{\Psi}\!\left(0\right)\!\right.\right\rangle$, the composition $\mathcal{T}_{-t}\!\circ\! \mathcal{T}_t$ will also be the identity morphism $\mathrm{id}^{2}_{\left|\!\left.{\Psi} \!\right.\right\rangle}$ on the Hilbert space of the system's state vectors $\left|\!\left.{\Psi} \!\right.\right\rangle$. But since for each vector $\left|\!\left.{\Psi} \!\right.\right\rangle$ the identity morphism is unique at any $t\geq 0$, $\mathrm{id}^{1}_{\left|\!\left.{\Psi} \!\right.\right\rangle}=\mathrm{id}^{2}_{\left|\!\left.{\Psi} \!\right.\right\rangle}$ and so\\

\begin{equation} \label{9} 
   \begin{array}{cl}
      t\geq 0:
      &
      \mathcal{T}_{-t}\!\circ\! \mathcal{T}_t
      =
      f\!\circ\! g 
      \;\;\;\;  .
   \end{array}  
\end{equation}
\smallskip

\noindent In this manner, one can conclude (by analogy to classical mechanics) that if quantum mechanics is a correct and complete theory, then the state vectors $\left|\!\left.{\Psi} \!\right.\right\rangle$ of a quantum system must be isomorphic to the elements $\mathpzc{Y}$ of the physical reality.\\

\noindent However, one must be warned that such a conclusion would be made without taking into account explicitly formulated steps necessary to construct the isomorphism $\mathpzc{Y}\simeq\left|\!\left.{\Psi} \!\right.\right\rangle$. Indeed, it is evident that in the real world to construct any specific morphism $h: A\!\rightarrow\! B$ (say, to identify an element $B$ if an element $A$ is given or to compute for each input $A$ a corresponding output $B$) would obviously take some time, say $T_h$. Furthermore, the more operations would be required to construct the morphism $h: A\!\rightarrow\! B$, the greater this time $T_h$ might be, meaning that $T_h$ must be a function of the number of such operations $N$.\\

\noindent Let $T_f(N)$ and $T_g(N)$ denote the upper bound of the amount of time required to construct the direct, $f: \mathpzc{Y}\!\!\rightarrow\!\! \left|\!\left.{\Psi} \!\right.\right\rangle$, and inverse, $g: \left|\!\left.{\Psi} \!\right.\right\rangle \!\!\rightarrow\!\! \mathpzc{Y}$, morphisms, respectively. In addition, let $T_{\mathcal{T}}(N)$ stand for the upper bound of the time needed to compute $\left|\!\left.{\Psi}\!\left(t\right)\!\right.\right\rangle$ at $t\!>\!0$ provided $\left|\!\left.{\Psi}\!\left(0\right)\!\right.\right\rangle$. Then, based on the equality (\ref{9}) one would get\\

\begin{equation} \label{10} 
   \begin{array}{cl}
      t > 0:
      &
      O\!\left(T_{\mathcal{T}}(N)+T_{-\mathcal{T}}(N)\right)
      =
      O\!\left(T_f(N)+T_g(N)\right)
      \ \ \mathrm{as} \, N \rightarrow \infty
      \;\;\;\;  ,
   \end{array}  
\end{equation}
\smallskip

\noindent where $T_{-\mathcal{T}}(N)$ is the upper bound of the time needed to compute $\left|\!\left.{\Psi}\!\left(0\right)\!\right.\right\rangle$ if $\left|\!\left.{\Psi}\!\left(t\right)\!\right.\right\rangle$ is provided at $t\!>\!0$. As the unitary operator $U\!\left(H(t),0,t\right)$ is symmetrical with respect to time, values $T_{\mathcal{T}}$ and $T_{-\mathcal{T}}$ should be equal; consequently, one finds the following equality\\

\begin{equation} \label{11} 
   \begin{array}{cl}
      t > 0:
      &
      O\!\left(T_{\mathcal{T}}(N)\right)
      =
      O\!\left(T_f(N)+T_g(N)\right)
      \ \ \mathrm{as} \, N \rightarrow \infty
      \;\;\;\;   
   \end{array}  
\end{equation}
\smallskip

\noindent required to hold true as a \textit{computational condition of the uniqueness of the identity morphism} on the Hilbert space of state vectors $\left|\!\left.{\Psi} \!\right.\right\rangle$. Clearly, if this condition could not be met, then at any $t\!>\!0$ there would be two distinguished by the time complexity identity morphisms $\mathrm{id}^{1}_{\left|\!\left.{\Psi} \!\right.\right\rangle}$  and $\mathrm{id}^{2}_{\left|\!\left.{\Psi} \!\right.\right\rangle}$  and thereby an isomorphism between states $\left|\!\left.{\Psi} \!\right.\right\rangle$ and elements of the reality $\mathpzc{Y}$ would be impossible.\\

\noindent Thus, the state vectors $\left|\!\left.{\Psi} \!\right.\right\rangle$  of a quantum system could be isomorphic to (i.e., in a one-to-one correspondence with) the ontic states $\mathpzc{Y}$ of this system (and accordingly support $\psi$-ontic models of the system) only if the computational condition (\ref{11}) were fulfilled.\\

\noindent The present paper will show that the condition (\ref{11}) cannot hold true for a generic physical system and hence the question of whether the state vectors $\left|\!\left.{\Psi} \!\right.\right\rangle$ are necessarily isomorphic to the elements of the reality cannot be resolved generally.\\

\section{Constructing morphisms between state vectors and elements of the reality}

\noindent Let us start by evaluating the upper bound $T_f (N)$ of the amount of time (or operations) required to construct the direct morphism $f: \mathpzc{Y}\!\!\rightarrow\!\! \left|\!\left.{\Psi} \!\right.\right\rangle$.\\

\noindent The procedure for assigning to any given physical state $\mathpzc{Y}$ of a system a corresponding state vector $\left|\!\left.{\Psi} \!\right.\right\rangle$ can be represented as a task of \textit{quantum state tomography}. Truly, quantum state tomography is the process of identifying the quantum state, say $\left|\!\left.{\Psi}\!\left(0\right)\!\right.\right\rangle$, of the system prepared in some well-characterized initial physical state, say $\mathpzc{Y}\!\left(0\right)$, by measurements on that system (actually, on a number of identical copies of that system each in the same physical state) – for a review of quantum tomography see \cite{Mohseni,Blume,Lvovsky}.\\

\noindent Let the initial physical state $\mathpzc{Y}\!\left(0\right)$ of the quantum system be characterized by a set of values of physical quantities $\mathpzc{Y}=\lbrace\mathpzc{y}_1,…,\mathpzc{y}_N \rbrace$, and let the probability of observing a particular value $\mathpzc{y}_i$ be $\Pr(\mathpzc{y}_i)$. According to the direct morphism  $f: \mathpzc{Y}\!\left(0\right) \!\!\rightarrow\!\! \left|\!\left.{\Psi}\!\left(0\right) \!\right.\right\rangle$, this probability $\Pr(\mathpzc{y}_i)$ should correspond to $\mathrm{Trace}\!\left(\mathpzc{y}_i \rho\right)$ where $\rho =\left|\!\left.{\Psi}\!\left(0\right) \!\right.\right\rangle \!\! \langle{\Psi}\!\left(0\right)\!\vert$ represents the density matrix of the system. Assuming that the measurement for each $\mathpzc{y}_i$ is repeated $C$ times and the value $\mathpzc{y}_i$ appears $c_i$ times, the probability $\Pr(\mathpzc{y}_i)$ can be estimated as $\frac{c_i}{C}$, therefore solving the following equation for each $\mathpzc{y}_i$\\

\begin{equation} \label{12} 
     \mathrm{Trace}\!\left(\mathpzc{y}_i \rho\right)
      =
      \frac{c_i}{C}
      \;\;\;\;   
\end{equation}
\smallskip

\noindent one can, in aggregate, reconstruct the state vector $\left|\!\left.{\Psi}\!\left(0\right) \!\right.\right\rangle$.\\

\noindent In the special case, where all physical quantities $\mathpzc{Y}=\lbrace\mathpzc{y}_1,…,\mathpzc{y}_N \rbrace$ of the quantum system can take in only two values, say $\pm$1, the density matrix $\rho$ can be written as\\

\begin{equation} \label{13} 
     \rho
      =
      \left|\!\left.{\psi}_1\right.\right\rangle \!\dots\! \left|\!\left.{\psi}_N\right.\right\rangle \!
      \langle{\psi}_N\vert  \!\dots\!  \langle{\psi}_1\vert 
      \;\;\;\;  , 
\end{equation}
\smallskip

\noindent where each $\left|\!\left.{\psi}_i\right.\right\rangle$ is

\begin{equation} \label{14} 
      \left|\!\left.{\psi}_i\right.\right\rangle
      =
      a_{i1}\!\left|\!\left.{\mathpzc{y}_i=+1}\!\right.\right\rangle
      + 
      a_{i2}\!\left|\!\left.{\mathpzc{y}_i=-1}\!\right.\right\rangle
      \;\;\;\;  , 
\end{equation}
\smallskip

\noindent (coefficients $a_{i1,2}$ are complex and ${\vert a_{i1}\vert}^2+{\vert a_{i2}\vert}^2=1$). This implies that the upper bound of the time required to construct the direct morphism $f: \mathpzc{Y}\!\left(0\right) \!\!\rightarrow\!\! \left|\!\left.{\Psi}\!\left(0\right) \!\right.\right\rangle$ would be $T_f (N)=O(4^N)$ since in this case one would need up to $4^N-1$ operations (such as projective measurements) to identify the quantum state vector $\left|\!\left.{\Psi}\!\left(0\right) \!\right.\right\rangle$ of the system.\\

\noindent The exponentiality of $T_f (N)$ can be significantly lowered if a quantum system can be easily prepared in a given initial physical state.\\

\noindent For example, the quantum system, whose all physical quantities $\mathpzc{Y}=\lbrace\mathpzc{y}_1,…,\mathpzc{y}_N \rbrace$ can take in only two values $\pm$1, can be realized as a system of $N$ spin-$\frac{1}{2}$ particles, where $\mathpzc{y}_i=+1$ corresponds to the $i^{\mathrm{th}}$ spin being, say, up in the $z$-direction and hence $\mathpzc{y}_i=-1$ corresponds to the $i^{\mathrm{th}}$  spin down in the $z$-direction. For such a $N$-qubit system, the expression (\ref{14}), in which $a_{i1,2}=1/\sqrt{2}$, will describe the equal superposition of the two computational basis states $\left|\!\left.{0}\!\right.\right\rangle$ and $\left|\!\left.{1}\!\right.\right\rangle$ corresponding to the $i^{\mathrm{th}}$ spin right-aligned in the $x$-direction. So, coupling a magnetic field in the $x$-direction to each of $N$ spin-$\frac{1}{2}$ particles, one can – in polynomial in $N$ number of operations – prepare the system in the physical state $\mathpzc{Y}$ that would correspond to the state vector $\left|\!\left.{\Psi}\!\left(0\right) \!\right.\right\rangle$\\

\begin{equation} \label{15} 
      \left|\!\left.{\Psi}\!\left(0\right) \!\right.\right\rangle
      =
      \left|\left.{x}_1=0\right.\right\rangle \!\dots\! \left|\left.{x}_N=0\right.\right\rangle \!
      \ \ \ 
      \mathrm{with} 
       \ \ \ 
      \left|\left.{x}_i=0\right.\right\rangle 
        =
      \frac{1}{\sqrt{2}}\!\left|\left.{{z}_i=0}\right.\right\rangle
      + 
      \frac{1}{\sqrt{2}}\!\left|\left.{{z}_i=1}\right.\right\rangle
    \;\;\;\;  . 
\end{equation}
\smallskip

\noindent Consequently, the upper bound of the amount of time (i.e., operations) required to construct the direct morphism $f: \mathpzc{Y}\!\left(0\right) \!\!\rightarrow\!\! \left|\!\left.{\Psi}\!\left(0\right) \!\right.\right\rangle$ in this case would be $T_f (N)=\mathrm{poly}(N)$.\\

\noindent As to the upper bound $T_g (N)$ of the amount of time required to construct the inverse morphism $g:  \left|\!\left.{\Psi}\!\right.\right\rangle\!\rightarrow\! \mathpzc{Y}$, it is easy to show that like $T_f (N)$ this value can be polynomial in $N$ too.\\

\noindent Suppose that at the time $t$ the quantum state $\left|\!\left.{\Psi}\!\left(t\right) \!\right.\right\rangle$ of the $N$-qubit system is described by the zero ground state $\left|\!\left.{\psi}_b\right.\right\rangle=\left|\left.{z}_1\right.\right\rangle \!\dots\! \left|\left.{z}_N\right.\right\rangle$ of the quantum version of the Ising Hamiltonian $H_{\mathrm{Ising}}(\mathpzc{y}_1,…,\mathpzc{y}_N)$\\

\begin{equation} \label{16} 
      H_{\mathrm{Ising}}(\mathpzc{y}_1,…,\mathpzc{y}_N)
      =
      \left(
      \sum^{N}_{i=1}{n_i\mathpzc{y}_i} 
      \right)^2
    \;\;\;\;  , 
\end{equation}
\smallskip

\noindent where $n_i$ are some positive numbers (for work discussing the particular choice of the Ising Hamiltonian see \cite{Fischer,Bapst,Lucas}). Then, measuring each physical quantity $\mathpzc{y}_i$ and substituting the measurement outcome  – either +1 or -1 – into the expression (\ref{16}) one can – in polynomial in $N$ number of operations – verify that $H_{\mathrm{Ising}}(\mathpzc{y}_1,…,\mathpzc{y}_N)=0$ and thus that the physical state $\mathpzc{Y}\!\left(t\right) $ of the $N$-qubit system is really in correspondence with the zero ground state $\left|\!\left.{\psi}_b\right.\right\rangle$ at the moment $t$, proving in that way that $T_g (N)=\mathrm{poly}(N)$.\\

\noindent As it follows from (\ref{11}), in the case where both values $T_f (N)$ and $T_g (N)$ are polynomial in $N$, the computational condition of the uniqueness of the identity morphism on the Hilbert space of state vectors $\left|\!\left.{\Psi} \!\right.\right\rangle$ at any $t>0$ turns into the requirement imposed on the upper bound $T_{\mathcal{T}}(N)$ of the time needed to compute the quantum state $\left|\!\left.{\Psi}\!\left(t\right) \!\right.\right\rangle$ at $t>0$ if the initial state $\left|\!\left.{\Psi}\!\left(0\right) \!\right.\right\rangle$ is given\\

\begin{equation} \label{17} 
   \begin{array}{cl}
      t > 0:
      &
      O\!\left(T_{\mathcal{T}}(N)\right)
      =
      \mathrm{poly}(N)
      \ \ \mathrm{as} \, N \rightarrow \infty
      \;\;\;\;   .
   \end{array}  
\end{equation}
\smallskip

\noindent Let us evaluate the feasibility of this requirement.\\

\section{Constructing the automorphism corresponding to adiabatic evolution}

\noindent With this purpose, consider the $N$-qubit system that at the initial point of time $t=0$ is described by the state vector $\left|\!\left.{\Psi}\!\left(0\right) \!\right.\right\rangle$ presented in (\ref{15}). Suppose that at the final point of time $t=T_{\mathcal{T}}(N)>0$ the quantum state of this system $\left|\!\left.{\Psi}\!\left(t\right) \!\right.\right\rangle$ found through the nontrivial automorphism $\mathcal{T}_t: \left|\!\left.{\Psi}\!\left(0\right)\!\right.\right\rangle \!\!\rightarrow\!\! \left|\!\left.{\Psi}\!\left(t\right)\!\right.\right\rangle$ is described by the zero ground state $\left|\!\left.{\psi}_b\right.\right\rangle$ of the quantum version of the Ising Hamiltonian $H_{\mathrm{Ising}}$ displayed in (\ref{16}).\\

\noindent Clearly, in the case of such a system both values $T_f (N)$ and $T_g (N)$ would be polynomial in the number of system's qubits $N$. Let us estimate then the value $T_{\mathcal{T}}(N)$ to see whether it can be polynomial in $N$ as well.\\

\noindent As it can be readily seen, the initial state vector $\left|\!\left.{\Psi}\!\left(0\right) \!\right.\right\rangle$ of the studied system is the ground state of the Hamiltonian $H_{\mathrm{trans}}(\hat{\sigma}^{x}_{1},\dots,\hat{\sigma}^{x}_{N})$ consisting of transverse magnetic fields\\

\begin{equation} \label{18} 
      H_{\mathrm{trans}}(\hat{\sigma}^{x}_{1},\dots,\hat{\sigma}^{x}_{N})
      =
      -
       \sum^{N}_{i=1}{\hat{\sigma}^{x}_{i}} 
     \;\;\;\;  , 
\end{equation}
\smallskip

\noindent where $\hat{\sigma}^{x}_{i}$ is a Pauli $2\times2$ matrix, whose eigenvector $\left|\left.{x}_i=0\right.\right\rangle$ corresponding to the eigenvalue +1 is shown in (\ref{15}).\\

\noindent Thus, at the initial time $t=0$ the Hamiltonian of this system $H(t)$ matches the Hamiltonian $H_{\mathrm{trans}}(\hat{\sigma}^{x}_{1},\dots,\hat{\sigma}^{x}_{N})$ and at the final moment of time $t=T_{\mathcal{T}}(N)$ the Hamiltonian $H(t)$ is equal to $H_{\mathrm{Ising}}(\hat{\sigma}^{z}_{1},\dots,\hat{\sigma}^{z}_{N})$, the quantum version of the Ising Hamiltonian $H_{\mathrm{Ising}}(\mathpzc{y}_1,…,\mathpzc{y}_N)$, where each $\mathpzc{y}_i$ is replaced by the Pauli matrix $\hat{\sigma}^{z}_{i}$. In order to implement an interpolation (say, a linear one) between the initial and final Hamiltonians, $H_{\mathrm{trans}}$ and $H_{\mathrm{Ising}}$, the system's Hamiltonian $H(t)$ should take the form\\

\begin{equation} \label{19} 
      H(s)
      =
      (1-s)
       H_{\mathrm{trans}}(\hat{\sigma}^{x}_{1},\dots,\hat{\sigma}^{x}_{N})
      +
     s
     H_{\mathrm{Ising}}(\hat{\sigma}^{z}_{1},\dots,\hat{\sigma}^{z}_{N})
     \;\;\;\;  , 
\end{equation}
\smallskip

\noindent where $s=t/(T_{\mathcal{T}}(N))\in[0,1]$ is a reduced time.\\

\noindent To ensure that the system remains, with high probability, in the instantaneous ground state of the Hamiltonian (\ref{19}) at all subsequent times $s$ (otherwise values $T_f (N)$ and $T_g (N)$ would not be polynomial in $N$ together) the evolution of the Hamiltonian $H(s)$ must be \textit{adiabatic}, i.e., slow enough. Thus, the question about the feasibility of the requirement $O\!\left(T_{\mathcal{T}}(N)\right)=\mathrm{poly}(N)$ (necessary for the uniqueness of the identity morphism $\mathrm{id}_{\left|\!\left.{\Psi} \!\right.\right\rangle}$ at the moment $s\ne0$) comes to be the question whether the evolution time $T_{\mathcal{T}}(N)$ satisfying a criterion for the adiabatic approximation is polynomial in $N$.\\

\noindent According to the adiabatic theorem (e.g., see \cite{Messiah} or \cite{Elgart}), a generic criterion for the adiabatic approximation, which guarantees that the quantum adiabatic algorithm executing the automorphism $\mathcal{T}_s: \left|\!\left.{\Psi}\!\left(0\right)\!\right.\right\rangle \!\!\rightarrow\!\! \left|\!\left.{\Psi}\!\left(1\right)\!\right.\right\rangle$ will find the solution $\left|\!\left.{\Psi}\!\left(1\right)\!\right.\right\rangle=\left|\!\left.{\psi}_b\right.\right\rangle$ at the end of the evolution $s$, can be formulated as\\

\begin{equation} \label{20} 
      T_{\mathcal{T}}(N)
      \gg
      O\!\left(
            \frac{1}{\Delta^{2}_{\mathrm{min}}}
      \right)
     \;\;\;\;  , 
\end{equation}
\smallskip

\noindent where $\Delta_{\mathrm{min}}$ denotes the minimum gap between the two lowest levels – i.e., instantaneous eigenvalues $E_1(s)$ and $E_0(s)$ – of the Hamiltonian $H(s)$\\

\begin{equation} \label{21} 
      \Delta_{\mathrm{min}}
      =
      \min_{s\in[0,1]}
      \left(
            E_1(s)-E_0(s)
      \right)
     \;\;\;\;  . 
\end{equation}
\smallskip

\noindent Whether the quantum adiabatic algorithm is efficient (i.e., takes a polynomial amount of time) in finding $\left|\!\left.{\psi}_b\right.\right\rangle$ is not known (in accordance with \cite{Aharonov}, for discussion see \cite{vanDam,Reichardt}). Even though in absence of special symmetries in the Hamiltonian $H(s)$ (and when the number of system's qubits $N$ is finite) true crossings between $E_1(s)$ and $E_0(s)$ are not expected, i.e., $\forall s \in [0,1]:  E_1(s)-E_0(s)>0$, levels $E_1(s)$ and $E_0(s)$ may still get extremely close with the system size $N\rightarrow \infty$ (for example, exponentially close such that $\Delta_{\mathrm{min}}\sim\exp(-cN)$ where $c>0$ is some constant).\\

\noindent On the other hand, let us recall that finding the solution $\left|\!\left.{\psi}_b\right.\right\rangle$ to the quantum version of the Ising Hamiltonian (\ref{16}) with $H_{\mathrm{Ising}}=0$, that is, finding a configuration of spins $\left|\!\left.{\psi}_b\right.\right\rangle=\left|\left.{z}_1\right.\right\rangle \!\dots\! \left|\left.{z}_N\right.\right\rangle$, for which the sum of the $n_i$ for the $\left|\!\left.{{z}_i=0}\!\right.\right\rangle$ spins is the same for the sum of the $n_i$ for the $\left|\!\left.{{z}_i=1}\!\right.\right\rangle$ spins, would give a solution to \textit{the number partitioning problem} known to be NP-complete \cite{Karp}. For that reason, finding the solution $\left|\!\left.{\psi}_b\right.\right\rangle$ in polynomial time $\mathrm{poly}(N)$ would entail the equality of the computational complexity classes NP and P generally believed to be improbable (see, e.g., \cite{Gasarch}). Hence, for the considered above adiabatic $N$-qubit system the requirement $O\!\left(T_{\mathcal{T}}(N)\right)=\mathrm{poly}(N)$ should be regarded as unachievable.\\

\noindent As a result, doing ‘nothing’ by applying the identity morphism $\mathrm{id}^{2}_{\left|\!\left.{\Psi} \!\right.\right\rangle}=\mathcal{T}_{-1}\!\circ\! \mathcal{T}_1$ to the state vector $\left|\!\left.{\Psi}\!\left(0\right)\!\right.\right\rangle$ of this system should take much greater time than doing the same ‘nothing’ with the identity morphism $\mathrm{id}^{1}_{\left|\!\left.{\Psi} \!\right.\right\rangle}=f\!\circ\! g$. Such an inequality would imply that the diagram, which connects four elements $\left|\!\left.{\Psi}\!\left(0\right)\!\right.\right\rangle$, $\left|\!\left.{\Psi}\!\left(1\right)\!\right.\right\rangle$, $\mathpzc{Y}\!\left(1\right)$ and, again, $\left|\!\left.{\Psi}\!\left(1\right)\!\right.\right\rangle$ using four morphisms $\mathcal{T}_{1}$, $g$, $f$, and $\mathcal{T}_{-1}$, would not be commutative: The same endpoints $\left|\!\left.{\Psi}\!\left(1\right)\!\right.\right\rangle$ would lead to different – in terms of time complexity – results by the compositions $\mathcal{T}_{1}\!\circ\! \mathcal{T}_{-1}$ and $f\!\circ\! g$. In other words, the triangle $\left|\!\left.{\Psi}\!\left(1\right)\!\right.\right\rangle \rightarrow \mathpzc{Y}\!\left(1\right) \rightarrow\left|\!\left.{\Psi}\!\left(1\right)\!\right.\right\rangle$ containing both a state vector and an element of the reality would not be commutative to the triangle $\left|\!\left.{\Psi}\!\left(1\right)\!\right.\right\rangle \rightarrow \left|\!\left.{\Psi}\!\left(0\right)\!\right.\right\rangle \rightarrow\left|\!\left.{\Psi}\!\left(1\right)\!\right.\right\rangle$ containing only state vectors, meaning that state vectors are not equivalent to elements of the reality.\\

\noindent Ergo, unless P$=$NP, the concept of isomorphism existing between state vectors and the elements of the reality would be unsuitable for the given quantum system.\\

\section{Concluding remarks}

\noindent It is reasonable to represent quantum mechanics as a category \textbf{Hilb} consisting of sets of Hilbert spaces and morphisms, bounded linear operators, between elements of Hilbert spaces, state vectors $\left|\!\left.{\Psi} \!\right.\right\rangle$. On the other hand, the entire objective world can itself be represented as a category, say \textbf{Y}, consisting of structured sets $\mathpzc{Y}=\lbrace\mathpzc{y}_1,…,\mathpzc{y}_N \rbrace$ of values of physical quantities and relations between them (similar to the developed in \cite{Doering08} ‘category of systems’, \textbf{Sys}, whose objects are the physical systems of interest and their interrelations). So, the question is, are these two categories, \textbf{Hilb} and \textbf{Y}, isomorphic?\\

\noindent The answer to this question is important since it could resolve the longstanding problem of the interpretation of the state vector $\left|\!\left.{\Psi} \!\right.\right\rangle$ in quantum mechanics (there has been a lot of interest in this subject recently – see, for example, papers \cite{Hardy,Colbeck,Pusey,Ghirardi13,Ghirardi14}, just to name a few).\\

\noindent Indeed, let us assume that \textbf{Hilb} is isomorphic to \textbf{Y}; then it would imply that state vectors $\left|\!\left.{\Psi}\!\right.\right\rangle$ can be considered the same as elements of the reality $\mathpzc{Y}$; in other words, it would imply that for all practical purposes $\left|\!\left.{\Psi}\!\right.\right\rangle$ and $\mathpzc{Y}$ are identical. But this can only be compatible with the $\psi$-ontic models of quantum mechanics, in which one can assert that the state vector $\left|\!\left.{\Psi}\!\right.\right\rangle$ is a real thing, that is, $\left|\!\left.{\Psi}\!\right.\right\rangle$ represents a state of reality rather than a state of knowledge. Consequently, if the categories \textbf{Hilb} and \textbf{Y} are isomorphic, then quantum theory deals with the objective world as directly as does classical mechanics.\\

\noindent In the present paper, it was argued that to tackle the issue of the isomorphism $\mathbf{Hilb}\simeq\mathbf{Y}$, one needs to complement the category-theoretic approach to quantum mechanics with the computational-complexity-theoretic considerations. Upon doing so, one gets the computational condition of the uniqueness of the identity morphism on the Hilbert spaces in \textbf{Hilb} restricting the isomorphism between \textbf{Hilb} and \textbf{Y}.\\

\noindent As it was demonstrated in the paper, from a computational-complexity-theoretic perspective, the hypothesis of isomorphism that exists between state vectors $\left|\!\left.{\Psi} \!\right.\right\rangle$ and the elements of the reality $\mathpzc{Y}=\lbrace\mathpzc{y}_1,…,\mathpzc{y}_N \rbrace$ at $t>0$ is expected to be unsuitable for the adiabatic $N$-qubit system.\\

\noindent Thus, it is logical to conclude that the question of whether quantum state vectors can be considered the same as elements of the reality cannot be resolved generally, that is, for any time and for any physical system.\\

\end{document}